\documentclass[
 rsi,
amsmath,amssymb,
reprint,%
]{revtex4-2}

\usepackage{graphicx}
\usepackage{dcolumn}
\usepackage{bm}

\usepackage[utf8]{inputenc}
\usepackage[T1]{fontenc}
\usepackage{mathptmx}

\usepackage{xcolor}

\begin{document}

	\preprint{AIP/123-QED}
	
	\title[Sample title]{Two-color Kerr microscopy of two-dimensional materials with sub-picosecond time resolution}
	
	\author{M. Kempf}
	\email{michael.kempf@uni-rostock.de}
	\author{A. Schubert}%
	\author{R. Schwartz}
	\author{T. Korn}
	\email{tobias.korn@uni-rostock.de}
	\affiliation{Institute of Physics, University of Rostock 
	}%

	\date{\today}
	
	\begin{abstract}
We present a two-color Kerr microscopy system based on two electronically synchronized erbium-fiber laser oscillators with independently tunable emission energies spanning most of the visible spectrum. Combining a spatial resolution below 2~$\mu$m and sub-ps time resolution with high sensitivity and cryogenic sample temperatures, it is ideally suited for studying spin and valley dynamics in a wide range of two-dimensional materials.  We illustrate its capabilities by studying a monolayer of the common semiconducting transition metal disulfide MoS$_2$.
	\end{abstract}
	
	\maketitle
		\section{\label{intro}Introduction}
Time-resolved Kerr rotation (TRKR), as well as  the related time-resolved Faraday rotation~\cite{Baumberg1994} (TRFR) and resonant spin amplification~\cite{PhysRevLett.80.4313} (RSA) are well-established techniques for studying spin dynamics in semiconductors. They have been applied in a variety of semiconducting materials~\cite{Kerr-CdTe,Kerr-GaN,RSA-ZnO}, their heterostructures~\cite{Harley-Kerr,  Kosaka2009,Korn2010415} and ensembles of nanostructures~\cite{PhysRevLett.96.227401,Holleitner_2007, PhysRevB.80.104436}. Frequently, tunable laser sources are employed in these techniques, as resonant pumping of an interband  transition typically yields a higher-fidelity optical orientation of spin polarization~\cite{H_bner_2008} and the Kerr rotation angle is enhanced around interband resonances. A prime example for a successful symbiosis of  laser sources and materials is the combination of Ti:sapphire laser oscillators and heterostructures based on gallium arsenide, whose resonances lie perfectly within the laser tuning range.

In recent years, two-dimensional (2D) materials have garnered a lot of research attention. The semiconducting transition metal dichalcogenides (TMDCs) are of particular interest due to their peculiar band structure, which leads to spin-valley locking~\cite{Xiao2012}, and this topic has spawned a lot of research activity  (see, e.g., Refs. \cite{Xu2014,Liu2019,Glazov2021} and references therein). The interband optical selections allow for generation of a combined spin and valley polarization via circularly polarized light  that can also be read out optically via helicity-resolved photoluminescence (PL). However, the large oscillator strengths of excitons in these materials lead to ultrashort radiative lifetimes~\cite{Poellmann2015}, so that some of the  dynamics are at the limits of time-resolved detection using, e.g., streak camera systems~\cite{Marie16}. 
Therefore, pump-probe-based techniques such as TRKR, whose time resolution is typically only limited by the laser pulse duration, are, in principle, ideally suited for studying spin and valley dynamics in TMDCs, and first studies using degenerate energies for pump and probe were performed several years ago~\cite{PhysRevB.90.161302,Yang2015a,Plechinger2016-Kerr}. 
However, degenerate pump-probe spectroscopy typically requires a spatial separation of pump and probe beam paths, e.g., via different incidence angles on the sample. Such an approach is not readily compatible with the use of high-aperture lenses or microscope objectives required for obtaining a spatial resolution on the order of a micrometer, and therefore, relatively large, homogeneous sample areas are needed when degenerate TRKR is employed. This provides a significant limitation especially in the field of 2D materials, where samples are often prepared in a nondeterministic way by exfoliation from bulk crystals, resulting in flakes with spatially inhomogeneous number of layers and typical lengths scales of a few microns. When two or more 2D materials are to be combined into a van der Waals (vdW) heterostructure (HS), this issue is exacerbated, as interlayer contact in a vdW HS is seldom homogeneous over more than a few square microns, even if the HS area itself can be made larger, and the relative alignment of the adjacent layers (twist angle) also varies locally~\cite{Parzefall_2021}. Additionally, the use of nondegenerate pump and probe energies may give access to relaxation or up-conversion dynamics~\cite{Liu-APL20}.

Here, we present a two-color Kerr microscopy system based on two electronically synchronized  laser oscillators which are independently tunable, allowing for nondegenerate TRKR and transient reflectivity measurements with collinear pump and probe beams coupled into a microscope objective. It offers a very large  tuning range in the visible spectrum that is energetically above the range accessible to Ti:sapphire laser oscillators, and incidentally covers the  A and B exciton transitions in the disulfide-based TMDCs MoS$_2$ and WS$_2$. It is also spectrally well-matched to many other material systems such as various lead halide perovskites~\cite{Belykh2019,Klein-perovskite}.  As a test sample, we utilize a monolayer of the prototypical TMDC MoS$_2$. We demonstrate the  versatility and sensitivity of our setup by performing wavelength-dependent TRKR and differential reflectance measurement series at low temperatures.
\section{\label{experiment}Experimental setup}
\begin{figure}
	\includegraphics[width=\linewidth]{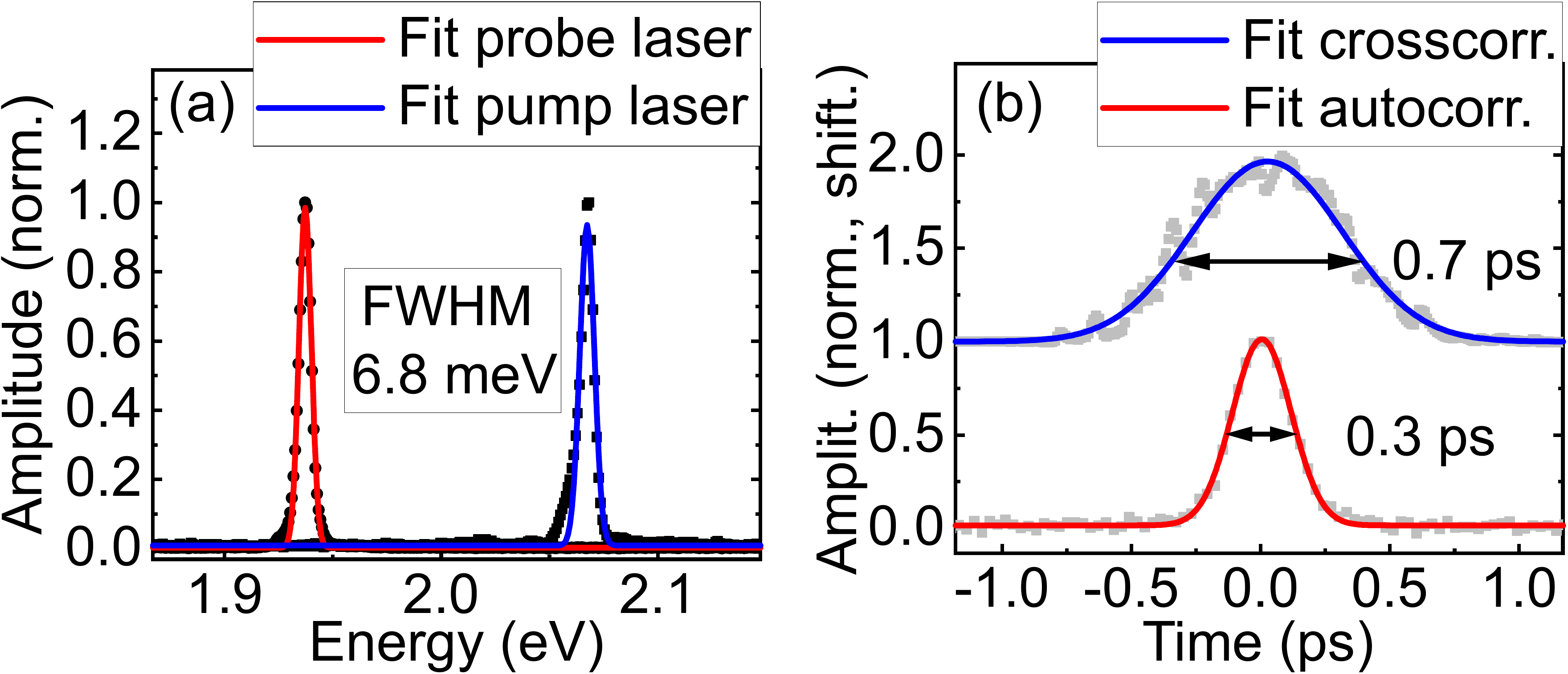}
	\caption{\label{LaserChar-2panel} (a) Spectra of pump and probe lasers measured with spectrometer integrated into the setup. The linewidths (full width at half maximum extracted from a Gaussian fit) are typically about 6.8~meV. (b) Autocorrelation (bottom trace) and cross-correlation (top trace) traces of the laser pulses. The cross-correlation trace was measured using the SHG crystal integrated into the setup, the autocorrelation was measured in a non-collinear geometry.}
\end{figure}

At the heart of our experiment is a pair of electronically synchronized laser oscillators based on erbium-fiber lasers. They contain a mode-locked master oscillator and a core-pumped high power amplifier. Mode locking is established via a saturable absorber mirror. The two laser oscillators are electronically synchronized by means of fast electronics. Synchronization of their repetition rates and phases is realized via control of the resonator length  of one of the oscillators. For this, the position of a mirror in the free-space region of the oscillator beam path is controlled by a piezo translation stage. 
The amplified laser pulses  are fed into a highly nonlinear fiber, and the supercontinuum generated within the fiber is frequency converted using a manually tunable second-harmonic generation (SHG) crystal, allowing for continuous wavelength tuning. Two motorized prism compressors allow adjustment of the supercontinuum and  optimization of the pulse width. Both lasers have a tuning range of about 490~nm to 700~nm (corresponding to 1.77~eV to 2.53~eV). The spectral tuning of the lasers requires optimizing prism compressors and thus changes the effective beam path length by introducing variable amounts of dispersive media. This induces a variable time delay between the trigger impulse extracted from the master oscillator that is used for electronic synchronization and the laser output. Therefore, the temporal overlap of the laser pulses from the synchronized lasers has to be confirmed, either via cross-correlation measurements, or directly via the pump-probe traces, whenever wavelength changes are made.
The spectral linewidth of the laser emission is typically about 6.8~meV (full width at half maximum (FWHM) extracted from a Gaussian fit), as Fig.~\ref{LaserChar-2panel}(a) shows. Autocorrelation measurements for an individual oscillator yield FWHM of about 0.3~ps, and cross-correlation measurements in our setup show FWHM on the order of 0.7~ps, indicating the low temporal jitter of the synchronization scheme. 
\begin{figure}
	\includegraphics[width=\linewidth]{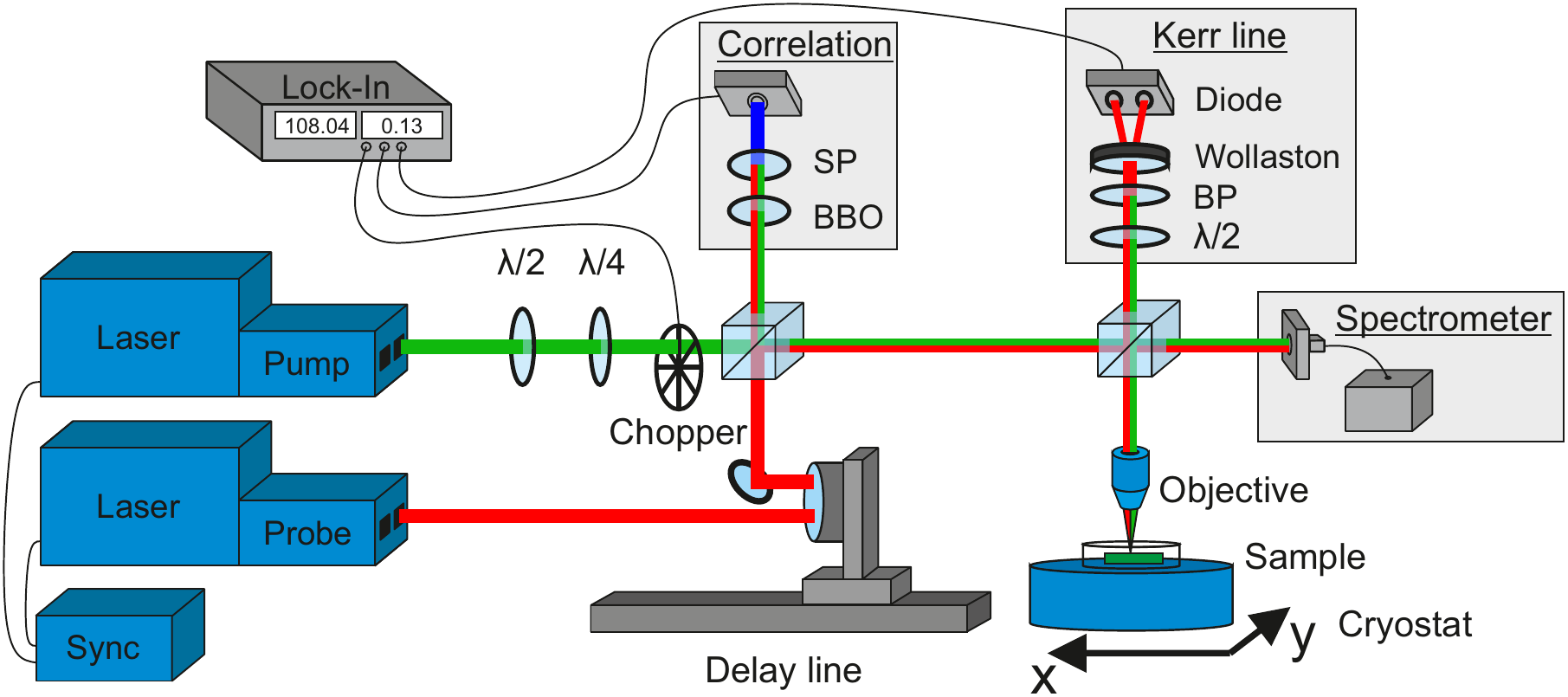}
	\caption{\label{Setup}Schematic of the experimental setup. The two lasers are synchronized electronically. The pump beam polarization state is defined using wave plates, and its amplitude is modulated by a flywheel chopper for lock-in detection. The probe beam path is guided over a mechanical delay stage to control pump-probe time delay. Subsequently, the two beams are combined at a beamsplitter cube and focused onto the sample mounted on the cold finger of a flow cryostat. The reflected probe beam is spectrally filtered using a band pass and its polarization state is analyzed using an optical bridge. Ancillary beam paths allow for simultaneous cross-correlation and spectral characterization of pump and probe beams.}
\end{figure}
The overall experimental scheme is depicted in Fig.~\ref{Setup}: time delay between pump and probe lasers is realized by a mechanical delay stage allowing for a maximum time delay of 4~ns. After the polarization of both beams is defined using achromatic wave plates and the pump beam passes a flywheel chopper, the two beams are made collinear using a nonpolarizing beam splitter cube and coupled into a small He-flow cryostat using a 80x microscope objective and an additional beam splitter cube. The focal spot size of the two beams is about 1.2~$\mu m$ (FWHM, determined by scanning over a lithographically defined metal structure). The reflected light is spectrally filtered using a band pass, which suppresses the pump beam. To facilitate measurement series in which the probe beam wavelength is varied over a large spectral range, a revolver-type mount for several band pass filters spanning the desired spectral range is integrated.  The pump-induced change of the reflected probe beam intensity (transient reflectivity $\Delta R$) or Kerr rotation $\Theta_K$ is detected via a photodiode ($\Delta R$) or an optical bridge ( $\Theta_K$) using a lock-in amplifier. The optical bridge consists of an achromatic half-wave plate and a Wollaston prism, which splits the optical beam paths of perpendicular polarizations. The split beams are detected by a pair of photodiodes (Thorlabs PDB210A), whose amplified difference signal is detected by the lock-in amplifier. Prior to Kerr rotation measurements, with the pump beam blocked, the half-wave plate in the optical bridge is rotated so that the two photodiode signals are identical, resulting in zero difference signal. 

 At the beam splitter cubes, the part of the collinear beams not guided along the path to the cryostat is utilized for cross-correlation measurements and spectral characterization of the laser emission, respectively.  For the cross-correlation measurements, the collinear beams are focused onto a Beta Barium Borate (BBO) crystal using a short-focal-length lens. The SHG radiation is spectrally separated from the fundamental with a short pass (SP) filter and detected with a photodiode and the lock-in amplifier. Spectral characterization of the lasers is performed using a small, self-contained spectrometer with fiber coupling (Ocean optics HR4000).
Additionally, a white-light source for wide-field illumination and a digital camera are integrated into the beam path (not shown in Fig.~\ref{Setup}), so that the laser focus can be adjusted to a specific sample position by moving the cryostat beneath the fixed beam path using translation stages.

\section{\label{results}Results}
\begin{figure}
	\includegraphics[width=\linewidth]{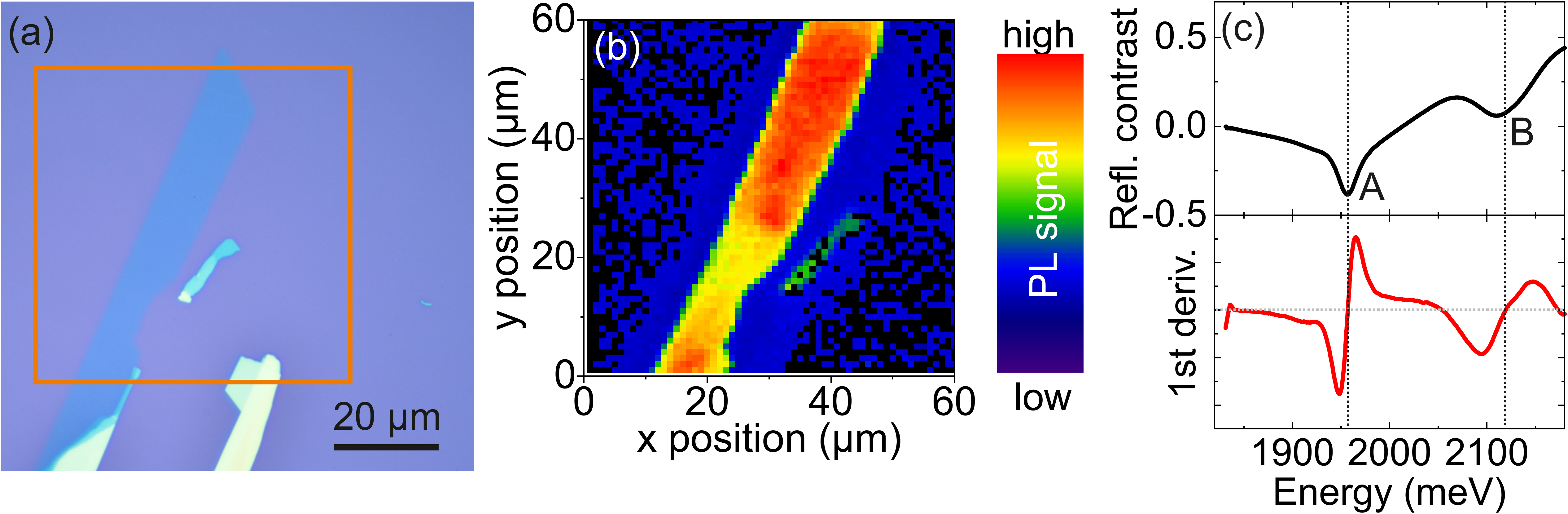}
	\caption{\label{MoS-3panel-PLscan-Whitelight}(a) Optical microscope image of large-area MoS$_2$ flake consisting predominantly of a monolayer region on top of Si/SiO$_2$ substrate. The orange square indicates the approximate area of the false-color PL map shown in (b). (b) False-color PL intensity map measured on flake shown in (a) at 80~K. (c) White-light reflectance contrast spectra (top) and first derivative of the spectra (bottom) measured on monolayer part of MoS$_2$ flake  at 80~K. The absorption features corresponding to A and B transitions are highlighted by  vertical dotted lines.}			
\end{figure}

As a test sample for our experiment, we prepared a  MoS$_2$ flake on a SiO$_2$-covered silicon substrate. Figure~\ref{MoS-3panel-PLscan-Whitelight}(a) shows an optical microscope image of the flake, which consists predominantly of a large monolayer region that is about 10 $~\mu$m wide and 80 $~\mu$m long. The monolayer region shows homogeneous PL intensity throughout, as the false color map in Fig.~\ref{MoS-3panel-PLscan-Whitelight}(b) indicates. Whitelight-reflectance contrast measurements reveal the characteristic absorption features associated with A and B exciton transitions.

All time-resolved measurements shown below were performed on the monolayer region of this flake. Let us first discuss the time resolution obtained in our measurements. 
	\begin{figure}
	\includegraphics[width=\linewidth]{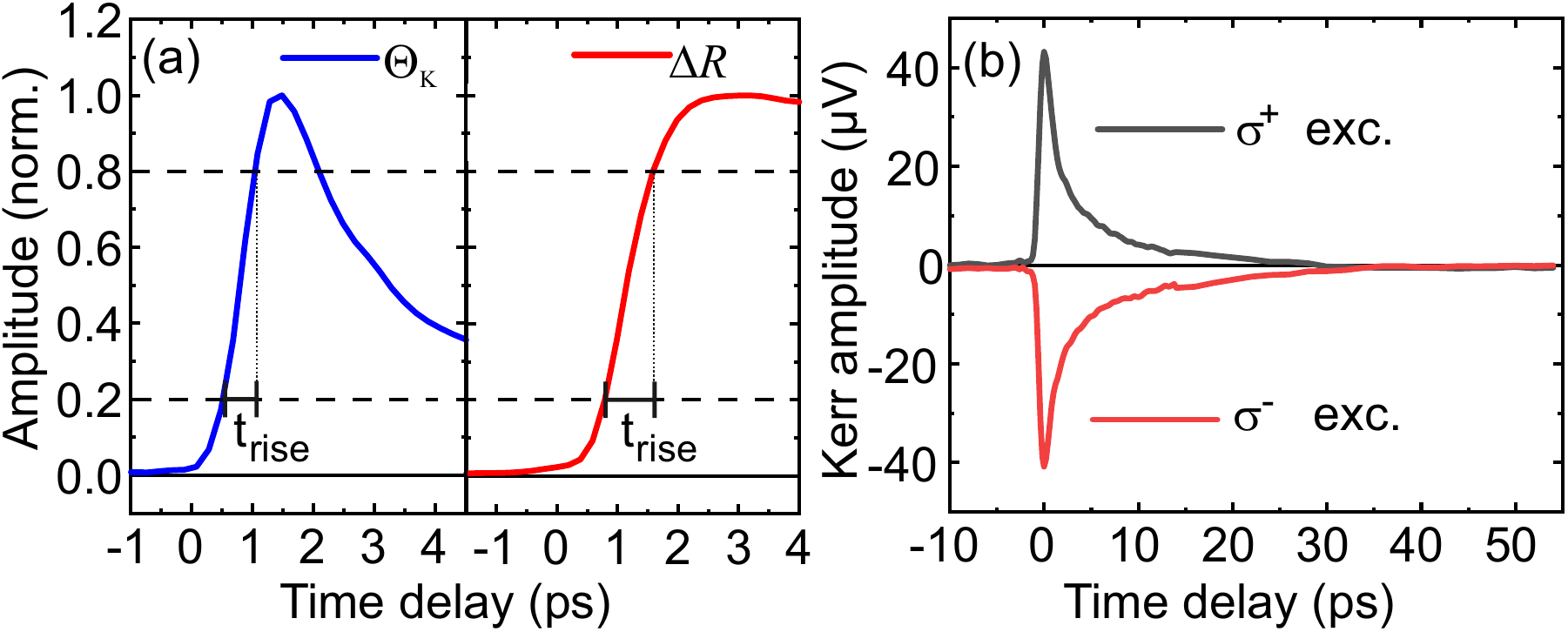}
	\caption{\label{Rise-Helicity-flip-2Panel}(a) Normalized high-resolution Kerr rotation (left panel) and differential reflectivity (right panel) traces measured on monolayer part of MoS$_2$ flake at 80~K. Risetimes defined via an increase of the signal from 20 to 80~percent of its maximum value are indicated. (b) Dependence of Kerr rotation signal on excitation helicity.}			
\end{figure}
Figure~\ref{Rise-Helicity-flip-2Panel}(a) shows typical normalized Kerr rotation and differential reflectivity measurements. For the differential reflectivity, we utilized a pump wavelength of 593~nm (close to the B exciton resonance) and a probe wavelength of 640~nm (close to the A exciton resonance), for Kerr rotation, a pump wavelength of 615~nm and a probe wavelength of 630~nm were used. We find risetimes (defined by an increase of the signal from 20 to 80 percent of its maximum value) between 0.5~ps and 0.8~ps, verifying that the optics in the beam path following the cross-correlation setup do not add significant  temporal broadening.  In order to suppress any signals not related to spin or valley polarization when measuring Kerr rotation, our experiment is set up to automatically perform two subsequent measurements with opposite excitation helicity (demonstrated in Fig.~\ref{Rise-Helicity-flip-2Panel}(b)). This is realized with a motorized rotator for the quarter-wave plate defining the pump beam polarization. By calculating the difference of the two resulting traces, any signal component that does not change its sign upon change of the excitation helicity is removed. 

The use of two independent laser oscillators allows us to arbitrarily tune either the pump or the probe beam energy to investigate resonance effects, while keeping the other energy fixed to realize identical pumping or probing conditions during such series. In monolayer MoS$_2$, the most prominent resonances that can be investigated are the A and B transitions. 

First, we perform a series of differential reflectivity measurements in which the pump beam energy is kept fixed slightly below the A exciton resonance, while the probe beam energy is varied in a broad spectral range encompassing the B exciton resonance. 
	\begin{figure}
	\includegraphics[width=\linewidth]{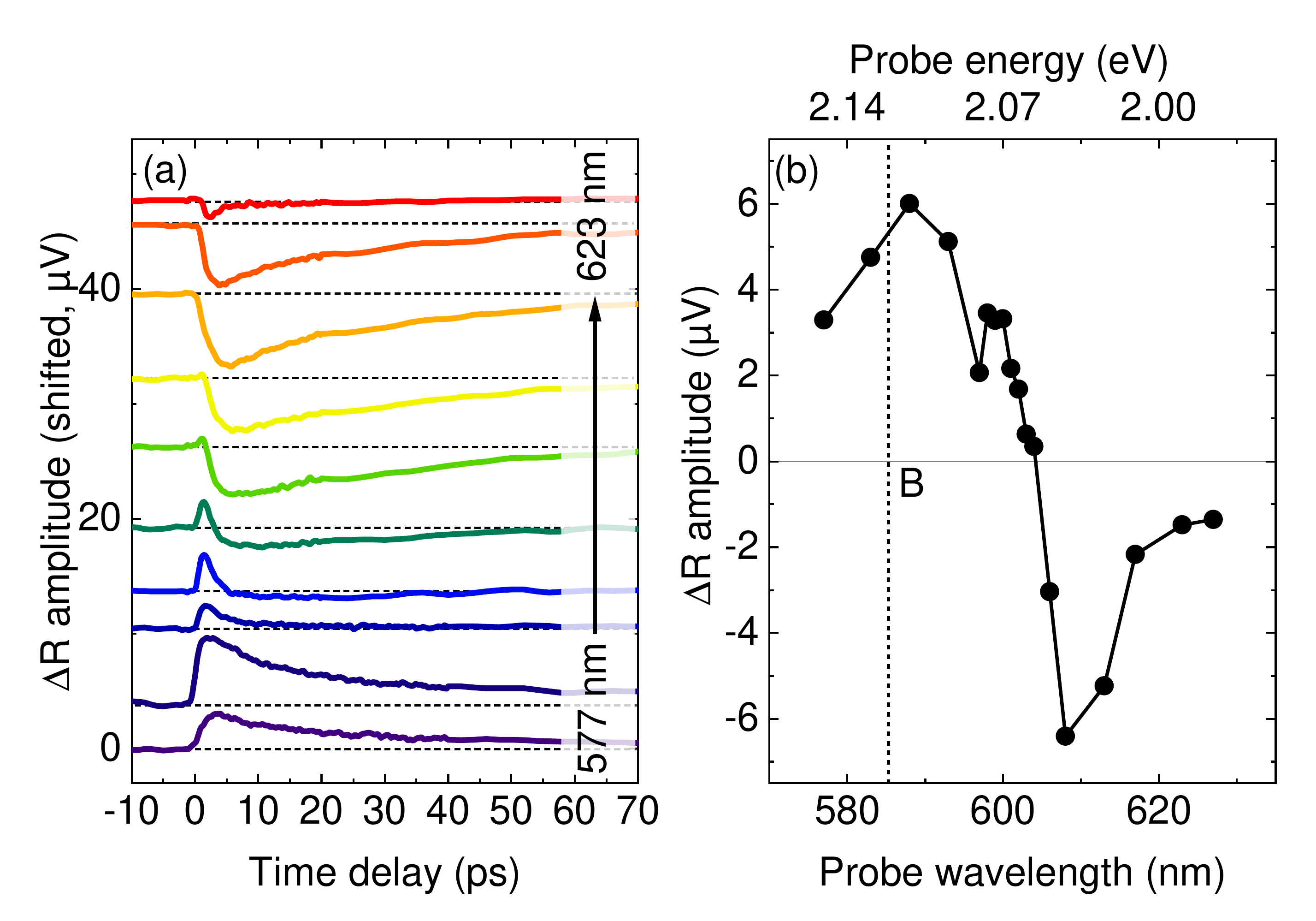}
	\caption{\label{DT-Probeseries-2Panel}(a) Differential reflectance traces measured on the monolayer part of a MoS$_2$ flake at 80~K using a fixed excitation wavelength for various probe wavelengths. (b) Dependence of maximum differential reflectance signal amplitude on probe wavelength extracted from a series of differential reflectance traces. The dotted vertical line marks the center of the B exciton resonance extracted from white-light reflectance measurements.}			
\end{figure}
Figure~\ref{DT-Probeseries-2Panel}(a) shows a representative part of such a series of $\Delta R$ traces measured at 80~K. We clearly see that for high probe beam energies (short wavelengths), the pump beam induces an increased reflectivity that decays on the order of about 15~ps, while for low probe beam energies (long wavelengths), a decreased reflectivity is induced that decays on similar timescales. In Fig.~\ref{DT-Probeseries-2Panel}(b), we plot the maximum induced reflectivity changes extracted from the whole measurement series, and  see that the maximum increase in reflectivity is close to the spectral position of the B exciton determined from our white-light reflectance measurements (Fig.~\ref{MoS-3panel-PLscan-Whitelight}(c)).
This probe-energy-dependent behavior corresponds to a pump-induced redshift of the B transition absorption feature, decreasing absorption at the original, unperturbed transition energy and increasing it at lower energy. Such a shift, induced by the optically generated carrier density, has been previously observed in various TMDCs using differential absorption and reflectivity measurements and attributed to band gap renormalization~\cite{Zhao-PumpProbe,Chernikov2015,Hanbicki-PumpProbe} or, more recently, to polaron formation~\cite{Voelzer21}. 

Next, we turn to the spectral dependence of Kerr rotation experiments. We perform a series of measurements with a constant pump energy and variation of the probe energy in a broad spectral range encompassing the A exciton resonance. 
	\begin{figure}
	\includegraphics[width=\linewidth]{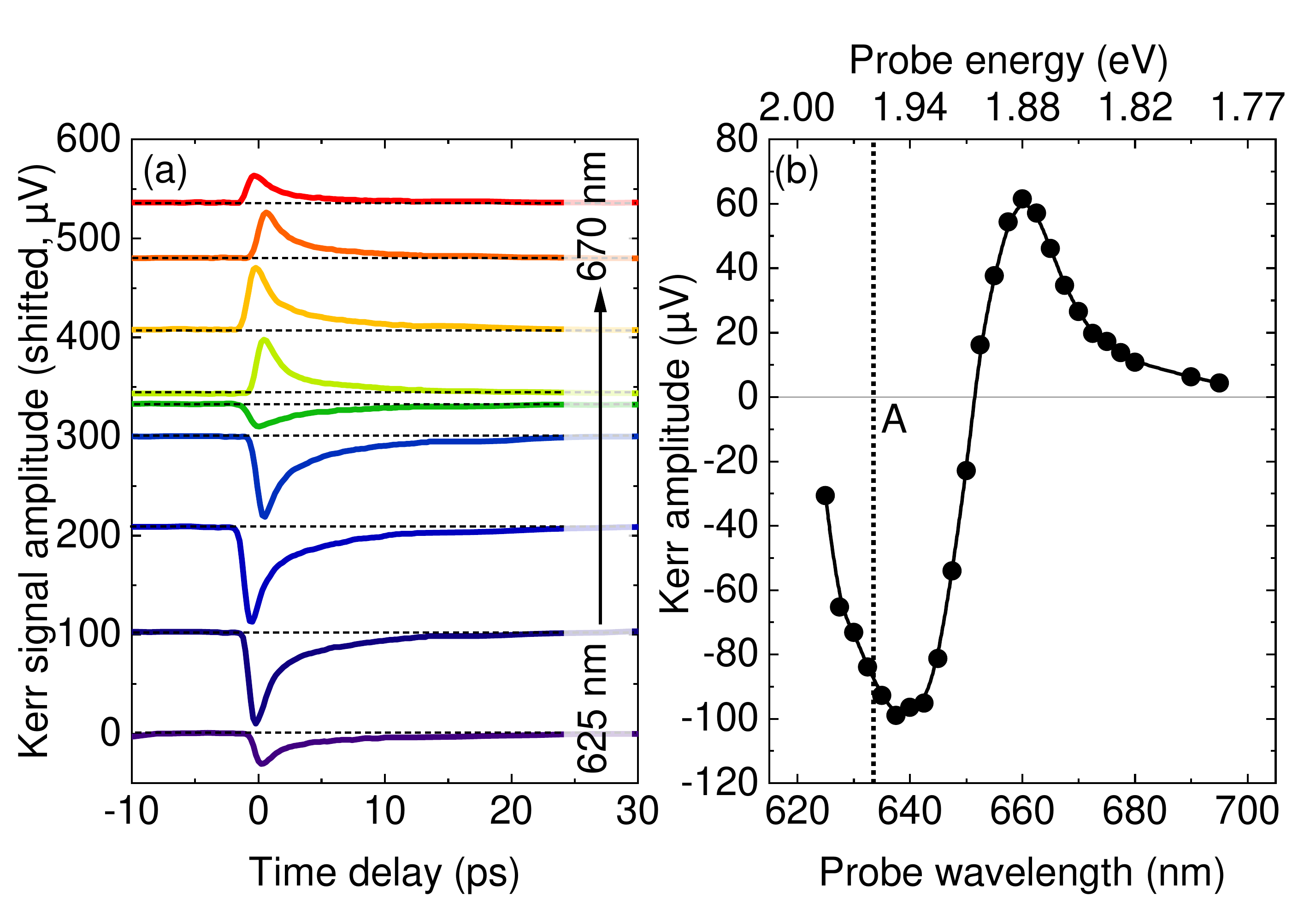}
	\caption{\label{Kerr-Probeseries-2Panel}(a) Kerr rotation traces measured on the monolayer part of a MoS$_2$ flake at 80~K using a fixed excitation wavelength for various probe wavelengths. (b) Dependence of maximum Kerr rotation signal amplitude on probe wavelength extracted from a series of Kerr traces.The dotted vertical line marks the center of the A exciton resonance extracted from white-light reflectance measurements.}			
\end{figure}
Figure~\ref{Kerr-Probeseries-2Panel}(a) shows a representative part of such a series of TRKR traces measured at 80~K. We clearly see that the amplitude of the Kerr signal strongly depends on the probe wavelength and even changes its sign. The maximum amplitudes extracted from this series are depicted in Fig.~\ref{Kerr-Probeseries-2Panel}(b) and clearly show the typical behavior of enhanced Kerr rotation around a resonance with its characteristic sign change. As indicated in Fig.~\ref{Kerr-Probeseries-2Panel}(b), however, the zero crossing is red-shifted with respect to the A exciton resonance extracted from white-light reflectance measurements. This behavior is to be expected, as the pump beam induces a redshift of  the A and B resonances due the excited photocarrier density as discussed above. 

We find that the dynamics observed in the TRKR traces are well-described using a biexponential decay function, in good agreement with a previous study using degenerate TRKR~\cite{Gerd-RRL}. We extract a time constant of about 8~ps for long-lived component of the biexponential decay, closely matching the previous results for a temperature of 80~K. In MoS$_2$, the dynamics of TRKR signal decay are driven by a combination of radiative recombination of excitons that are scattered into the light cone and exciton valley depolarization driven by long-range exchange interaction~\cite{Glazov2014}.

Next, we perform measurement series with fixed probe energy and systematically vary the pump energy.  
\begin{figure}
	\includegraphics[width=\linewidth]{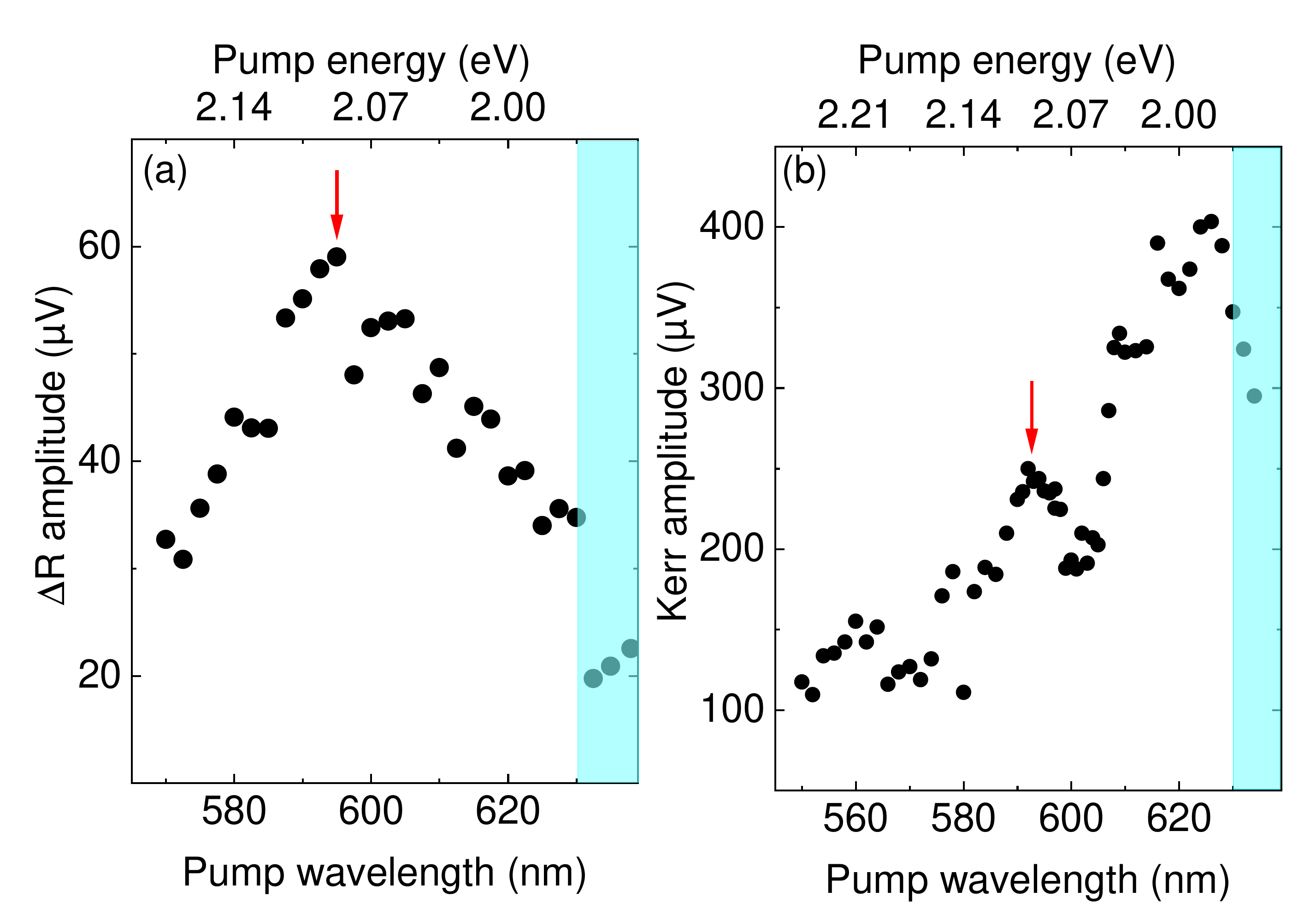}
	\caption{\label{PumpSeries-2Panel}(a) Dependence of maximum differential reflectance signal amplitude on pump wavelength extracted from a series of differential reflectance traces measured at 80~K. The red arrow marks the approximate position of the maximum at about 595~nm. (b) Dependence of maximum Kerr rotation signal amplitude on pump wavelength extracted from a series of Kerr traces measured at 4.2~K.The red arrow marks the approximate position of a local maximum at about 595~nm. In both panels, the blue shaded region marks the onset of the band pass filter in the detection beam path.}			
\end{figure}
Figure~\ref{PumpSeries-2Panel} shows the dependence of the maximum differential reflectance and Kerr rotation amplitudes on the pump energy. In both measurement series, the probe energy was kept fixed slightly below the A exciton resonance, and an appropriate band pass filter with a central wavelength of 640~nm was used to filter out the pump beam from the detection beam path. Close to its passband onset (marked by the blue-shaded area), the $\Delta R$ and Kerr signals are influenced by pump beam leakage.  
In the differential reflectance series, we observe a broad maximum of the $\Delta R$ amplitude at about 595~nm (2.08~eV), slightly below the B exciton resonance extracted from white-light reflectance measurements, corresponding to a resonantly enhanced absorption and increased reflectance change. By contrast, the maximum Kerr amplitude (Fig.~\ref{PumpSeries-2Panel}(b)) systematically increases as the pump energy is decreased towards the A exciton resonance. This behavior indicates that the fidelity of the valley polarization generation is reduced with increasing excess energy, most likely due to intervalley scattering during energy relaxation on timescales below the resolution of our experiment. In a similar vein, a systematic drop of photoluminescence circular polarization with increasing excess energy of the circularly polarized excitation was observed by several groups~\cite{Hanbicki-Excess,Lagarde-Excess,Maultzsch-Excess}.  Superimposed on this systematic behavior is a resonance at about 595~nm, corresponding to enhanced absorption as observed in Fig.~\ref{PumpSeries-2Panel}(a).    

\begin{figure}
	\includegraphics[width=\linewidth]{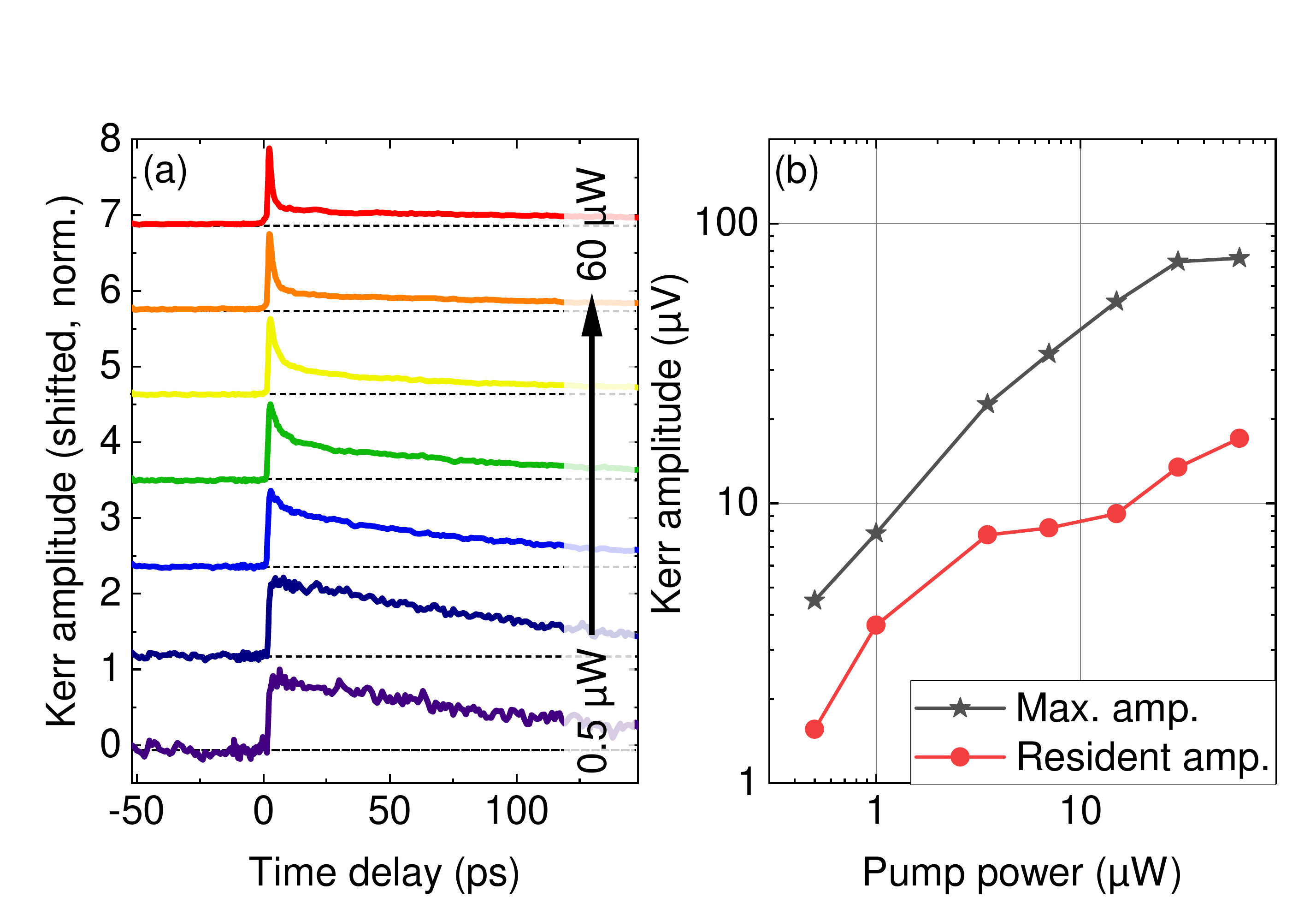}
	\caption{\label{Kerr-PowerSeries-2Panel}(a) Normalized Kerr rotation traces measured on the monolayer part of a MoS$_2$ flake at 4.2~K using various pump power levels. (b) Dependence of maximum Kerr rotation signal amplitude on pump power extracted from a series of Kerr traces. The black and red symbols refer to the maximum amplitude at zero time delay (black stars) and at a fixed, finite time delay of 100~ps (red dots).}			
\end{figure}

Finally, we vary the pump power. Figure~\ref{Kerr-PowerSeries-2Panel}(a) shows a series of normalized Kerr traces measured at 4.2~K, in which the pump power was varied over two orders of magnitude, while the probe power (50~$\mu$W), as well as pump (585~nm) and probe (640~nm) wavelengths, were kept fixed. A clear change of the Kerr trace shape can be observed. While low-power excitation yields a decay that can be described with a single exponential function and a decay constant of about 150~ps, medium- and high-power excitation yields a more complex behavior that has to be described using a biexponential decay, with a fast component  decaying in about 3~ps and a slow component with a decay constant of about 110~ps. The relative contribution of the slowly decaying component to the total signal systematically decreases with increasing pump power. In order to quantify this power dependence, we determined the maximum Kerr signal amplitude at zero time delay and at a fixed time delay of 100~ps, the data is depicted in Fig.~\ref{Kerr-PowerSeries-2Panel}(b). We clearly see that the maximum amplitude at zero time delay linearly increases over most of the power range investigated, with an onset of saturation at the highest pump power. By contrast, the amplitude at fixed time delay rapidly saturates with pump power after an initial linear increase.   
We can understand this behavior in the following way: the maximum Kerr amplitude at zero time delay is proportional to the density of valley-polarized excitons generated by the pump pulse, which scales linearly with the pump power until a bleaching of the absorption occurs, which we only reach for the highest pump fluences in the measurement series. By contrast, the Kerr signal at a time delay of 100~ps, which significantly exceeds the effective photocarrier lifetime at low temperatures~\cite{Korn_MoS,Lagarde-Excess,Marie16}, must stem from \emph{resident}, valley-polarized carriers that are present in the MoS$_2$ flake due to unintentional, inherent or absorbate-induced doping. A transfer of valley polarization from excitons to resident carriers had been observed in previous studies~\cite{Yang2015a,Schwemmer-resident-APL,Crooker-gated-PRL}. The resident carrier density naturally provides an upper limit for the valley polarization that can be generated in this way, and we are able to approach this upper limit at a pump power of about 10~$\mu$W for our particular sample. 
	\section{\label{conclusion}Conclusion and outlook}
In conclusion, we have demonstrated a system for two-color differential reflection and Kerr microscopy with combined spatial resolution below 2~$\mu$m and sub-picosecond time resolution in the visible range. We have validated its performance by studying  a monolayer of the TMDC MoS$_2$ at low temperatures. We were able to resonantly probe and/or excite photocarrier and spin-valley dynamics in this sample in a wide spectral range.
Our system is ideally suited for studying other 2D and layered materials, such as perovskites, as well as vdW HS consisting of different TMDCs, where it will give access to interlayer spin and valley polarization transfer between individual layers.

\begin{acknowledgments}
We are thankful for technical support by K. Kunze and C. Braun (Toptica) and fruitful discussion with M. Selig and A. Knorr. Funding by the DFG via SFB1277 (project B05) as well as grants KO3612/4-1 and KO3612/5-1 is gratefully acknowledged.
\end{acknowledgments}
\section*{DATA AVAILABILITY}
The data that support the findings of this study are available from the corresponding author 
upon reasonable request.	
		\appendix
\section{Sample preparation and characterization}
The MoS$_2$ flake was prepared using mechanical exfoliation from a commercially bought (HQ graphene) bulk MoS$_2$ crystal. It was initially prepared on top of an intermediate polydimethylsiloxane (PDMS) substrate and then transferred onto a silicon wafer covered with a 290~nm SiO$_2$ layer using a dry deterministic transfer process~\cite{Castellanos2014}. Low-temperature PL and reflectance contrast measurements were performed in a self-built microscope system using a 50x objective. For PL, a continuous-wave laser (532~nm wavelength) was focused onto the sample. For reflectance contrast, the emission from an intensity-stabilized tungsten lamp was collimated and focused onto the sample. The PL emitted from the sample, or the reflected white light, was collected by the same microscope objective,  dispersed in a grating spectrometer and detected using a Peltier-cooled charge-coupled-device (CCD) camera. The sample was mounted on the cold finger of a small cryostat, which can be scanned automatically beneath the fixed microscope beam path. For reflectance contrast measurements, a reference spectrum was recorded on the bare silicon/SiO$_2$ substrate, and the reflectance contrast $R$ was calculated as $R=\frac{I_{MoS_2}-I_{Ref}}{I_{Ref}}$, where $I_{MoS_2}$ and $I_{Ref}$ are the  intensities reflected from the MoS$_2$ flake and the bare substrate, respectively.
	\end{document}